# Early Detection of Long Term Evaluation Criteria in Online Controlled Experiments

Yoni Schamroth *, Liron Gat Kahlon * , Boris Rabinovich *, Professor David Steinberg ^

{yoni.schamroth@gmail.com, lirongat@gmail.com, borisrabin@gmail.com, dms@post.tau.ac.il}

*Perion

^Tel Aviv University

**Abstract.** A common dilemma encountered by many upon implementing an optimization method or experiment, whether it be a reinforcement learning algorithm, or A/B testing, is deciding on what metric to optimize for. Very often short-term metrics, which are easier to measure are chosen over long term metrics which have undesirable time considerations and often a more complex calculation. In this paper, we argue the importance of choosing a metrics that focuses on long term effects. With this comes the necessity in the ability to measure significant differences between groups relatively early. We present here an efficient methodology for early detection of lifetime differences between groups based on bootstrap hypothesis testing of the lifetime forecast of the response. We present an application of this method in the domain of online advertising and we argue that approach not only allows one to focus on the ultimate metric of importance but also provides a means of accelerating the testing period.

## 1. Introduction

### 1.1. Background

The use of optimization techniques in online media and marketing have become a crucial component in every aspect of the customer lifecycle for any website or mobile app. From customer acquisition and engagement to monetization of users, optimization algorithms provide the competitive edge necessary to stay afloat in an already crowded landscape. In each of these channels however, one of the biggest dilemmas often faced is what metric to optimize for. We argue here the importance of choosing a metric that focuses on long term effects over an extended period of time. We present a methodology enabling to us to measure and detect differences between test groups relatively early. The methodology presented in this paper is relevant for any optimization method of choice, but we will focus here primarily on controlled experimentation in the realm of the mobile and web environments. In this context, such experiments are commonly referred to A/B testing.

## 1.2. Controlled Experimentation

Controlled experimentation provides a relatively straightforward method for scientifically proving causality based on clear quantitative results. It is therefore finds itself at the cornerstone of multiple disciplines and is an indispensable tool for proving a hypothesis. With the recent explosion of data availability, we are seeing an increased demand for such experimentation, which not only uncovers vast amounts of knowledge lying dormant, but very often also results in a direct increase in a company's yield. The concept is very intuitive. We randomly assign two sets of users that can be regarded as being statistically equivalent with all but one feature. If these groups are found to differ significantly with regards to our OEC, or overall Evaluation Criteria, we can infer that the cause for the difference is related to the feature that has been altered. For a detailed description of best practices in running controlled experiments on the web see Kohavi et. al [9].

## 1.3. Choosing the correct OEC

The selection of the correct OEC or overall evaluation criteria is a topic that is unfortunately neglected by many. Choosing the correct OEC is often not a trivial task and requires both a broad understanding of the business and a clear agreement on what is the ultimate objective or objectives of the experiment should be. This metric can therefore vary from measures related to engagement, retention or revenue.

In the context of user acquisition, a good understanding of the purchase funnel is required. We therefore review some of the main stages during a potential user's lifecycle. The user passes through multiple stages leading up to a purchase (or ultimate action). These stages together with their corresponding metrics are presented below in the Figure 1 and Table 1, respectively.

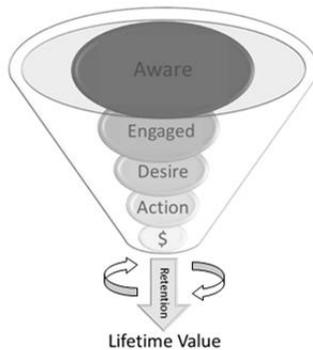

| **Stage** | **Metric** |
|---|---|
| Awareness | Impression or Mouse Over |
| Engagement | Page Views, Click through Rate |
| Desire | Session Duration; Likes |
| Action | Sign-up; Download; Install |
| Purchase | Sales count or Revenue |

Fig. 1: Purchase Funnel        Table 1: Stages and corresponding metrics within the Purchase Funnel.

The first step is to achieve awareness, this occurs when the user views the message or advertisement you are showing him. If we succeeded in attracting the user's interest, we establish engagement which represents the user's attention or interaction with what we are showing him. We can further measure his level of interest or desire by looking at the time he is spending on the site or in the case of an article the amount he is sharing it with others. The user may then perform some action which has significant meaning in itself or as a step closer to making a purchase. Finally, the user performs and purchase and money is transferred.

### 1.4. Life time metrics

On top of this funnel we can further calculate metrics which provide us with measures of customer loyalty. For example, metrics involving retention may describe the amount of time a user will stay loyal to one's product or what % of users will still remain active after a certain period.

When choosing an OEC it's important to focus on the ultimate KPI that is important to your business. This is generally the metric that is found at the end of the purchase funnel and, in most cases, is a measure related to revenue. However, it is imperative not to stop there. Very often companies ignore the long-term effect of retention which results in repeated actions or purchases thus increasing the user's true overall value[11] .

Metrics which encompass the effect of the entire funnel are ones that focus of long term effects. These generally include lifetime value, lifetime actions or lifespan. This is a measure of the overall contribution the user will generate over his lifetime (Novo 2001) or the duration of his lifespan itself. As we have witnessed many times and will demonstrate below, choosing to focus on short-term effects and ignoring the lifetime impact may lead to in incorrect conclusions – resulting in severe losses or missed opportunities.

### 2. Motivating Example and Objective

Below is an example, Figures 2-3, of one of the A/B tests conducted at Perion to assess the effectiveness of placing display advertisement on the search homepage. The hypothesis being examined: would the additional revenues being generated by the display ads surpass the expected loss in revenue stemming from users performing less search? The OEC chosen in this particular experiment was total revenue per user after a two-week period.

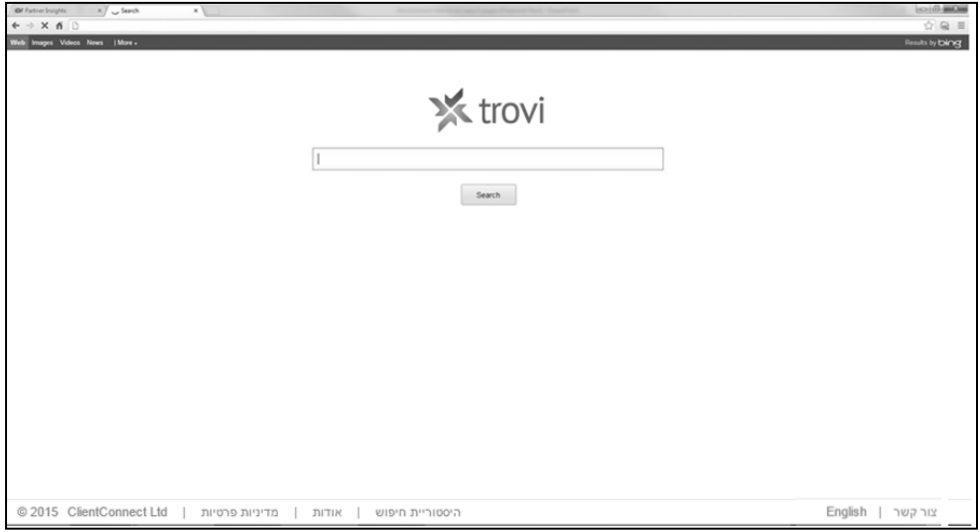

Figure 2: Control group of display advertisement experiment

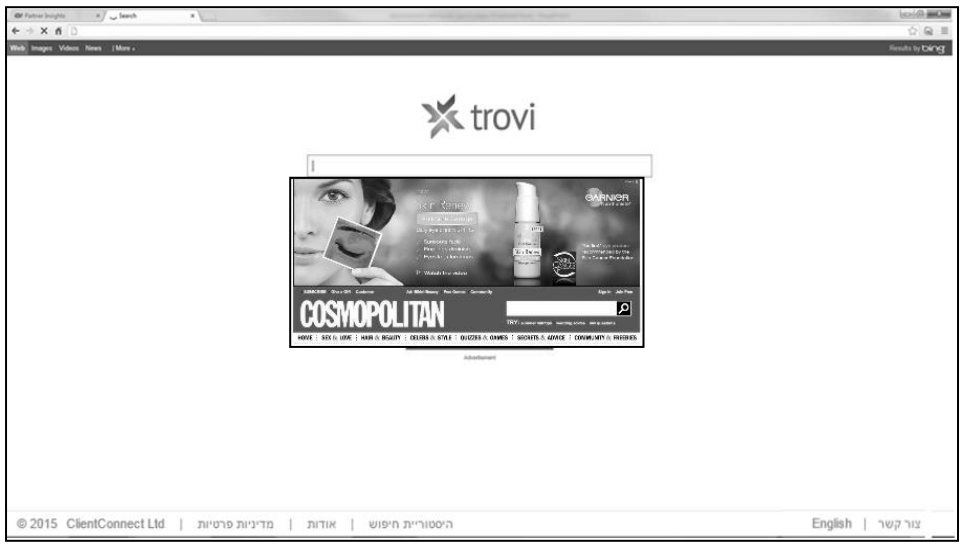

Figure 3: Test group of display advertisement experiment

The results can be seen in the Table 2 bellow:

|         | Users  | 14 Days | 28 Days | 60 Days | 90 Days |
|---------|--------|---------|---------|---------|---------|
| Test    | 98,344 | $0.278  | $0.397  | $0.483  | $0.538  |
| Control | 98,996 | $0.272  | $0.394  | $0.488  | $0.549  |
| **% Diff** |     | 2.20%   | 0.80%   | -1.00%  | -2.00%  |

Table 2: Experiment results where crossover was clearly evident

As can be seen, after a two week period the test showed a significant increase of 2.2% in net revenue per user in favor of the test group. The results were published, the winner declared and the development team started to implement display advertising on our homepage. However, we decided to continue to observe those users already recruited while they retained the configuration of their respective groups. The results were surprising. As time passed the overall revenue per user slowly shifted in favor of the control and eventually, after 3 months it was very clear that the control group was indeed the victor.

What was really happening is that users, upon first seeing the advertisement, were initially clicking resulting in an initial boost in revenues. However overall user satisfaction to the new layout actually decreased resulting in an increase in churn which in the long run resulted in an overall drop in revenues.

The example above clearly demonstrates the importance of focusing on long term effects. Technically however, this would require us to wait a lengthy period of time until we could finalize the outcome. The question is if we are able to reach a conclusion about long term differences early on. Looking at a chart of the relative difference between the groups in the experiment above, Figures 4.We can see that the trend is pretty clear and could have been forecasted earlier on.

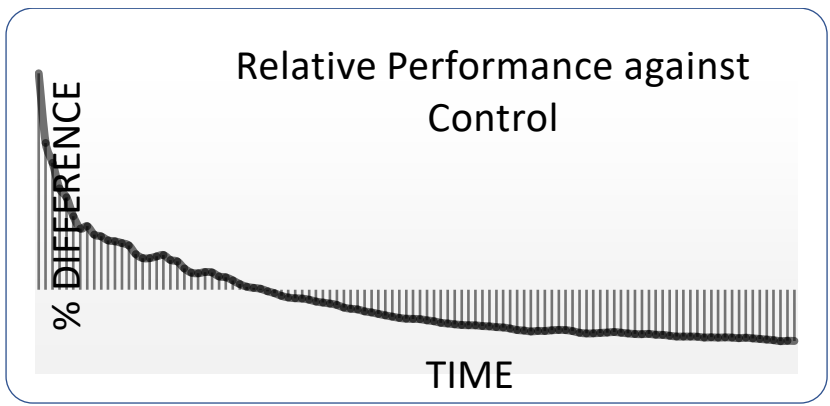

Figure 4: Relative performance of test group versus the control group

We provide here a method based on bootstrapping which enables us to quantify the uncertainty of the difference between groups in terms of lifetime metrics. To our knowledge the method we propose here is the first time such an issue is being addressed in an online marketing testing environment.

It has the following desirable properties in that (i) It allows for a relatively quick estimation, avoiding the necessity to maintain tracking of each test group for extended periods of time which can be limited. (ii) This method can be used in any testing or optimization environment, such that it overcomes the limitation in A/B testing mentioned by Nielson (2005) (see related works) (iii) It is non parametric - no assumptions are made as to the shape of the final distribution of the extrapolated response. (iv) No prior historic information is necessary, nor does it base itself on any prior assumptions of previous user behavior. This is can be very useful in cases which often occur where, due to tracking limitations, the response variable, say revenue, cannot be attributed directly to an end user but only to the test group itself. In such cases not only would we not be able to rely on a model of lifetime value at the user level, but further, since the recruitment period, rate and timing might vary from experiment to experiment, reliance on previously known behavior at the test group level also becomes challenging.

It is also important to note that in this research we are less concerned with our ability to forecast the actual lifetime value accurately but rather to quantify the probability of there being significant differences between the various test group expected values.

## 3. Related works

In a Monte Carlo study Mudelsee et. al [10] use bootstrapping techniques to quantify differences in two-sample environmental experiments. They further studied the robustness of the results in variations of the distributional shape of the data in terms of skewness and deviation from a standard Gaussian distribution. Hall and Wilson [6] highlight the importance of ensuring that resampling is done from within the assumptions of the null hypothesis when conducting non-parametric bootstrap hypothesis testing.

Nielsen [11] in his article on "Putting A/B Testing in its place", describes the necessity to focus on long-term behavioral metrics such as customer lifetime value. He lists this as a serious limitation of using A/B testing methodology. He argues that since, practically the ability to track each group over such a lengthy period is challenging, this proves to be a serious limitation to A/B testing methodology in general. This point is argued by Kohavi et.al [9] in their paper on controlled experiments on the web who agree on the importance e of measuring long term effects and suggest incorporating such a measurement within the chosen OEC of an A/B test.

In order to infer the effectiveness of a marketing campaign Brodersen etc. al [4] propose using Bayesian Structural Time-Series methodology in order to model the behavior of the response prior to intervention and then predict ahead a synthetic control simulating what would have happened had no intervention took place. Using this method one could estimate the difference in the response in what actually occurred in comparison to a forecasted control group.

Rosset et. al.[13] in their paper of" Customer Lifetime Value Models for Decision Support" review the importance of this metric and how it is used in the Telecommunications industry to measure the effectiveness of marketing campaigns. They present a segment-based model and discuss its implementation at the BI department at Amdocs.

Estimation of a products lifespan is an important metric in the piping domain. Plastic pipes for water and gas must provide long term durability for 50 years or higher. Since testing products for such long period is impractical, extrapolation techniques are justifiably applied to estimate lifetime differences at various service temperatures. Hoàng & Lowe [8] in their work used an empirical model based on the Arrhenius fit to estimate the extrapolated durability of the PE100 water pipe under various conditions. Suk, et al. [1], similarly used a Weibull-Arrhenius model to extrapolate the lifetime durability of Direct Methanol Fuel Cells (DMFC) thereby accelerating the time required for degradation tests.

## 5. Method

Our approach is to employ resampling techniques to estimate the distribution of the forecasted response of each test group individually. Once this is achieved we are then able to calculate a bootstrapped P-Value of the difference between the test group and control.

The bootstrap procedure originally proposed by [2] allows us to approximate an unknown distribution by using an empirical distribution obtained by resampling. The technique of bootstrapping was later extended to regression models to help approximate the distribution of the coefficients and prediction errors [5]

In the context of measuring lifetime differences within an A/B test, we propose treating each test group independently. Consider a response variable for group A say $Y_1 \ldots Y_n$ for the last n time intervals $T_1 \ldots T_n$. Though any best fitting model can be applied, specifically to the data in this study we used a log-log regression model to fit the response as a function of time. That's is:

$$\ln(Y_i) \sim \beta_0 + \beta_1 \ln(T_i) + \varepsilon_i$$

With $\varepsilon_i$ IID $N(0, \sigma^2)$. If necessary one can also incorporate other confounding variables such as weekday etc. We estimate the residuals by $\hat{\varepsilon}_i = Y_i - \hat{Y}_i$ where $\hat{Y}_i$ is the expected value of $Y_i$ obtained from the original regression. We repeat the next steps many times. In each iteration we sample n data points $\hat{\varepsilon}_1^* \ldots \hat{\varepsilon}_n^*$ with replacement from the empirical distribution of $\hat{\varepsilon}_i$. We then use this sample to generate pseudo-data $Y_i^*$ as follows $Y_i^* = \hat{Y}_i + \hat{\varepsilon}_i^*$. We then re-model the pseudo response and use the model to extrapolate ahead into the future to complete say 365-time intervals. The lifetime value estimate LTV for iteration j is then calculated as:

$$LTV_j = \sum_{i=1}^{365} Y_{ij}^*$$

Using this method, we are able to obtain an estimate of the distribution of the forecasted life time value for each group. This can then be used to obtain a P-Value as a measure of significance for the difference between the means of the groups.

**Our proposed solution can be summarized in the following steps:**

For each test group:
1. Fit a least squared regression on the full sample dependent on time and where necessary other variables.
2. Calculate residuals $\hat{\varepsilon}_i = Y_i - \hat{Y}_i$ where $Y_i$ represents 'Average Revenue' on day i
3. Draw a bootstrap random sample on size N from the distribution of residuals with replacement
4. Compute bootstrap values $Y_i^*$ by adding resampled residuals to the original regression fit
5. Fit a least squared regression on the bootstrap sample dependent on time
6. Use this regression to extrapolate x time units ahead and aggregate to obtain forecast of Life Time Value (LTV)
7. Repeat steps 3 to 6 many times to obtain extrapolated distribution of Life Time Value

Calculate the distribution of the difference in life time value of the control versus the test group.
The bootstrapped P-Value is then the proportion of differences greater than zero.

## 6. Result

In a retrospective study we applied the proposed approach over a variety of Perion internal experiments where lifetime metrics were of interest. Different types of lifetime KPI's were considered including lifetime searches and lifetime revenue. There were a total of 123 experiments that were re-evaluated.

Each experiment was re-analyzed in a similar manner. After a 14-day period into each test we evaluated the winner, once using the proposed method described above and a second time using a standard parametric testing procedure, equivalent to a t-test, not taking into account future trends. Success was measured after 60 days for each experiment. The decision policy that was used was to always choose the control group unless there was significant evidence that the test group was better. The results are summarized in Table 3.

|  | Success | Failure | Accuracy |
|---|---|---|---|
| Standard Test Accuracy | 112 | 11 | 91% |
| Proposed Test Accuracy | 119 | 4 | 97% |

Table 3: Relative performance of test group versus the control group

Overall our approach achieved 97% accuracy in determining the winning group versus 91% where standard test procedure where performed based on short term metrics. Taking future values into account would have resulted in different conclusions approximately 6% of the time. The differences stemmed from two different scenarios: cases where there was no significant evidence of a clear winner using standard techniques whereas taking a trend into account discovered that there actually was a significant difference and cases where there was an actual crossover, such that one group was significant early on but as time passed the results were reversed as in our motivating example.

## 7. Summary and Conclusions

We have attempted to illustrate in this article the importance of focusing on long term effects when deciding upon the right key performance indicator that is relevant to one's business. This has two immediate ramifications. The first is that one must be aware of the powerful influence of a time trend effecting the results. This trending effect not only could point to differences that, though now not significant, will clearly be significant in the future, but also could point to a possibility of the results crossing over in the future.

The second ramification, once we have established the need to account for this trend, is the ability to detect these long-term effects early on. To this end we propose using bootstrapping techniques which will allow us to attain a measure of statistical significance as to the future difference between the two groups. Most importantly, the proposed method would provide a method of making conclusion regarding the winner of an A/B test at early stages of the experiment, thus saving lots of time and money.